\documentclass[aps,prb,eqsecnum]{revtex4}

\usepackage{epsf}
\usepackage{amsfonts,amssymb,amsmath}

\newcommand{\be}{\begin{equation}}
\newcommand{\ee}{\end{equation}}
\newcommand{\bea}{\begin{eqnarray}}
\newcommand{\eea}{\end{eqnarray}}

\def \la{\langle}
\def \ra{\rangle}

\begin{document}

\title{Investigation of surface critical behavior of 
semi-infinite systems with cubic anisotropy.}
\author{Z. Usatenko}

\affiliation{
Institute of Physical Chemistry of the Polish Academy of Sciences,
Warsaw, 01-224, Poland }

\affiliation{
Institute for Condensed Matter Physics of the National 
Academy of Sciences of Ukraine Lviv, 79011, Ukraine }

\email[E-mail address:]{pylyp@icmp.lviv.ua}

\date{\today}

\begin{abstract}

The critical behavior at the special surface transition and crossover 
behavior from special to ordinary surface transition in semi-infinite 
$n$-component anisotropic cubic models are investigated by applying the 
field theoretic approach directly in $d=3$ dimensions up to the two-loop 
approximation. The crossover behavior for random semi-infinite 
Ising-like system, which is the nontrivial particular case of the cubic 
model in the limit $n\to 0$, is also investigated. 
 The numerical estimates of the resulting 
two-loop series expansions for the critical exponents of 
the special surface transition, surface crossover critical exponent 
$\Phi$ and the surface critical exponents of the layer, $\alpha_{1}$, and 
local specific heats, $\alpha_{11}$, are computed by means of Pad\'e and 
Pad\'e-Borel resummation techniques. For $n<n_{c}$ the system belongs to the 
universality class of the isotropic $n$-component model, while for $n>n_{c}$
the cubic fixed point is stable, where $n_{c}$ is the marginal spin 
dimensionality of the cubic model. The obtained results 
indicate that the surface critical behavior of semi-infinite systems 
with cubic anisotropy is characterized by new set of surface critical 
exponents  for $n>n_{c}$. 
 
\end{abstract}

\maketitle

\renewcommand{\theequation}{\arabic{section}.\arabic{equation}}
\section{Introduction}
\setcounter{equation}{0}

The investigation of the critical behavior of real systems is a very 
important task of the condensed matter theory. The critical 
behavior in such systems as polymers, easy-axis ferromagnets, 
superconductors, as well as superfluid $^{4}{\mathrm He}$, Heisenberg 
ferromagnets and quark-gluon plasma is described by isotropic $O(n)$ model 
with $n=0,1,2,3$ and $4$, respectively, and is very well determined in the 
framework of different theoretical and numerical approaches. 

Investigation of the critical behavior of real cubic crystals is 
the subject of extensive theoretical work during past decades. In real 
crystals, due to their complex crystalline structure,  some 
kind of anisotropy is always present. The simplest nontrivial crystalline 
anisotropy is a cubic one. A most typical model for description of the 
critical behavior of such systems is the $O(n)$ model with a cubic 
interaction term $\frac{v_{0}}{4 !}\sum_{i=1}^{n}\phi_{i}^{4}$ added to the 
usual $\frac{u_{0}}{4 !}(\mid \vec{\phi}\mid^{2})^{2})$ term  
\cite{Ahar73,KW73,W73}. This $n$-component cubic model is the particular 
case of an $mn$-component model with cubic anisotropy at $m=1$.
 Anisotropic $n$-component cubic model exhibits different types 
of continuous and first-order phase transitions, depending on the number of 
spin components $n$, space dimensionality $d$, and sign of the cubic 
coupling constant $v_{0}$. Cubic models are widely applied to the study of 
magnetic and structural phase transitions. In the limiting case of $n\to 0$, 
 it describes the critical behavior of random Ising-like systems. The 
case $m=1$ and $n\to \infty$ corresponds to the Ising model with equilibrium 
 magnetic impurities \cite{Aharony73}. Depending on the sign of the cubic 
 coupling constant $v_{0}$, two types of order are possible: along the 
 diagonals of a hypercube in $n$ dimensions $(v_{0}>0)$ or along the easy 
 axes $(v_{0}<0)$. 
In the latter case the system can undergo the first order phase transition, 
as was confirmed in some experiments \cite{Sznajd84}.
In the present work we are concerned with the case $v_{0}>0$. 

The presence of surfaces, which are inevitable in real systems, leads to  
additional complications. General reviews on surface 
critical phenomena are given in Refs. \cite{B83,D86,D97}. 
A typical model to study critical phenomena in real physical systems 
restricted by a single planar surface is the semi-infinite model \cite{B83}. 
There are different surface universality classes, defining the critical 
behavior in the vicinity of boundaries, at temperatures close to the bulk 
critical point ($\tau=(T-T_{c})/T_{c}\to 0$). Each bulk universality class, 
in general, divides into several distinct surface universality classes. 
Three surface universality classes, called ordinary ($c_{0}\to \infty$), 
special ($c_{0}=c_{sp}^{*}$) and extraordinary ($c_{0}\to -\infty$), are 
relevant for our case \cite{D86,D97,DD81}; they meet at a multicritical 
point $(m_{0}^{2},c_{0})=(m_{0c}^{2},c_{sp}^{*})$, which  corresponds 
to the special transition and is called the special point. The couplings 
$m_{0}$ and $c_{0}$ are defined in Eq.(\ref{2}) below.

Theory of critical behavior of individual surface universality classes 
is very well developed for pure isotropic systems 
\cite{D97,DD81,DD83,DDE83,DC91,DSh98}, the systems with quenched 
surface-enhancement disorder \cite{DN90,PS98,D98} and the systems with a 
random quenched bulk disorder at the ordinary and the special surface 
transitions \cite{UShH00,UH02}. General irrelevance-relevance criteria of 
the Harris type for the systems with quenched short-range correlated 
surface-bond disorder were predicted in \cite{DN90} and confirmed by 
Monte-Carlo calculations \cite{PS98,ILSS98}. Moreover, it was established 
that the surface critical behavior of semi-infinite systems with quenched 
bulk disorder is characterized by the new set of surface critical exponents 
in comparison with the case of pure systems \cite{UShH00,UH02}.

Experimental systems are typically characterized by the parameters different 
from the fixed point values. Investigations of the crossover behavior 
between different surface universality classes, however, have been limited 
to pure, isotropic systems, in which quite rich and nontrivial crossover 
behavior has been found \cite{DC91,DSh98,BL83,RCz96,CR97,MCD99}.
In the anisotropic systems the crossover behavior can be even more complex, 
because when the rotational symmetry is broken, the orientation of the 
surface may affect the surface critical phenomena \cite{Di}, and this effect 
can be different for different surface universality classes. 

In the present work we study 
the critical behavior at the special surface transition 
and crossover behavior from the special to the ordinary surface transitions, 
occuring in semi-infinite $n$-component model with cubic anisotropy.
 
The effective Landau-Ginzburg-Wilson Hamiltonian of the model under 
consideration in the semi-infinite space is given by 

\begin{eqnarray} 
H(\vec{\phi}) & = &\int_{0}^\infty dz \int d^{d-1}r [\frac{1}{2} 
\mid \nabla\vec{\phi} \mid ^{2} +  
\frac{1}{2} m_{0}^{2}\mid \vec{\phi} \mid^{2}\nonumber\\
&+& \frac{1}{4!} v_{0} \sum_{i=1}^{n} \phi_{i}^{4} + \frac{1}{4!}u_{0} 
(\mid \vec{\phi}\mid^{2})^{2})] + \int d^{d-1}r  
\frac{1}{2} c_{0}\vec{\phi}^{2},\label{2} 
\end{eqnarray}  
where $\vec{\phi}(x)$ is an $n$-vector field with the 
components $\phi_{i}(x)$, $i=1,...,n$.
Here $m_{0}^2$ is the "bare mass", representing a linear measure of the 
temperature difference from the critical point value. The parameters 
$u_{0}$ and $v_{0}$ are the usual "bare" coupling constants $u_{0} > 0$ and 
$v_{0} > 0$. In the case of replica limit $n\to 0$, the Hamiltonian 
(\ref{2})  describes the critical behavior of semi-infinite Ising-like 
systems with a random quenched bulk disorder, when $u_{0} < 0$ and $v_{0} > 
0$ (see \cite{UShH00,UH02,UH03}). The constant $c_{0}$ is 
related to the surface enhancement, which measures the enhancement of the 
interactions at the surface. It should be mentioned that the 
$d$-dimensional spatial integration is extended over a half-space 
$I\!\!R^d_+\equiv\{{\bf x}{=}({\bf r},z)\in I\!\!R^d\mid {\bf r}\in 
I\!\!R^{d-1}, z\ge 0\}$, bounded by a planar free surface at $z=0$. 
The fields $\phi_{i}({\bf r},z)$ 
satisfy the Dirichlet boundary 
condition in the case of ordinary transition: $\phi_{i}({\bf r},z)=0$ at 
$z=0$ and the Neumann boundary condition in the case of 
special transition:    
$\partial_{z}\phi_{i}({\bf r}, z)=0$ at $z=0$ 
\cite{DD81,DDE83}. The model defined in (\ref{2}) is translationally 
invariant in directions parallel to the external surface, $z=0$. Thus, we 
shall use mixed representation, i.e. Fourier representation 
in $d-1$ dimensions and real-space representation in $z$ direction. 

The added cubic term breaks the $O(n)$ invariance of the model, leaving a 
discrete cubic symmetry. The model (\ref{2}) has four fixed points:  
the trivial Gaussian, the Ising one in which $n$ components are decoupled, 
the isotropic ($O(n)$-symmetric) and the cubic fixed points. The Gaussian 
and Ising fixed points are never stable for any number of components $n$. 
For isotropic systems, the $O(n)$-symmetric fixed point is 
stable for $n<n_{c}$, whereas for $n>n_{c}$ it becomes unstable. 
Here $n_{c}$ is the marginal spin dimensionality of the cubic model, at 
which the isotropic and cubic fixed points change stability, i.e. for 
$n>n_{c}$, the cubic fixed point becomes stable. The $O(n)$-symmetric fixed 
poit is tricritical. At $n=n_{c}$, the two fixed points should coincide, and 
logarithmic corrections to the $O(n)$-symmetric critical exponents are 
present. The calculation of the critical marginal spin dimensionality 
$n_{c}$ is the crucial point in studing the critical behavior in 
three-dimensional cubic crystals. Different results for $n_{c}$ have been 
published in a series of works in which different methods have been used.
 In the framework of the 
field-theoretical RG analysis the one-loop and three-loop approximations at 
$\epsilon=1$ lead to the conclusion that $n_{c}$ should lie between 3 and 4 
\cite{KetleyW73,AharonyBruce74}, and the cubic ferromagnets are described 
by the Heisenberg model. 
On the other hand, by using the field theoretic 
approach directly in $d=3$ dimensions up to the three-loop approximation, it 
has been found that $n_{c}=2.9$ \cite{MayerSokolov87,Sh89}. 
Similar conclusions were obtained in Ref.\cite{YalabicHoughton77}, where 
it was found that $n_{c}=2.3$. The calculations performed by Newman and 
Riedel \cite{NR82} with the help of the scaling-field method, developed by 
Goldner and Ridel for Wilson's exact momentum-space RG equations, have given 
for $d=3$ the value $n_{c}=3.4$. Field-theoretical analysis, based on the 
four-loop series in three dimensions \cite{MSSh89,V00} and results of the 
five-loop $\epsilon=4-d$ expansion \cite{V00,KSch95,ShAS97} suggest that 
$n_{c}\le 3$. Recently a very precise six-loop result for the 
marginal spin dimensionality of the cubic model, $n_{c}=2.89(4)$, was 
obtained in the framework of the 3D field-theoretic approach \cite{CPV00}. 
Thus, it was finally established that the critical behavior of the cubic 
ferromagnets is not described by the isotropic Heisenberg Hamiltonian, but 
by the cubic model, at the cubic fixed point. However, it was found that 
the difference between the values of the bulk critical exponents at the 
cubic and the isotropic fixed points is very small, and it is hard to 
measure this difference experimentally. The recently obtained results 
stimulate us to perform the investigation of the {\it surface critical 
behavior} of semi-infinite n-component anisotropic cubic model, and to 
determine corresponding surface critical exponents.

The calculations are performed by applying the field theoretic approach 
directly in $d=3$ dimensions, up to the two-loop order approximation. The 
numerical estimates of the resulting two-loop series expansions for 
the critical exponents of the special surface transition, and for the 
surface crossover exponent $\Phi$ from the special to the ordinary 
transition and surface critical exponents of the layer, $\alpha_{1}$, and 
the local specific heat, $\alpha_{11}$, are computed by means of the Pad\'e 
\cite{B75} and Pad\'e-Borel \cite{BNGM76} resummation techniques for the 
cases $n=3,4,8$ and for the case of $n\to \infty$, which corresponds to the 
Ising model with equilibrium magnetic impurities.
The crossover behavior from the special to the ordinary transition for 
random semi-infinite Ising-like system, which is the nontrivial particular 
case of the cubic model in the limit $n\to 0$, is also investigated. 

\renewcommand{\theequation}{\arabic{section}.\arabic{equation}}
\section{Renormalization}
\setcounter{equation}{0}

In order to investigate the critical behavior at the special 
surface transition in the semi-infinite $n$-component anisotropic cubic 
model and to calculate the crossover exponent $\Phi$, we should consider 
correlation functions with insertions of the surface operator $\phi_{s}^2$,

\be
G^{(N,M;L_{1})}(\{{\bf x}_{i}\},\{{\bf r}_{j}\},\{{\bf R}_{l}\})=\la 
\prod_{i=1}^{N} \phi({\bf x}_{i})\prod_{j=1}^{M}\phi_{s}({\bf r}_{j}) 
\prod_{l=1}^{L_{1}}\frac{1}{2}\phi_{s}^{2}({\bf R}_{l})\ra, \label{7} 
\ee
which involve $N$ fields $\phi({\bf{x}}_{i})$ at distinct points 
${\bf{x}}_{i}(1\leq i \leq N)$ off the surface, $M$ fields 
$\phi({\bf{r}}_{j},z=0)\equiv \phi_{s}({\bf{r}}_{j})$ at distinct surface 
points with parallel coordinates ${\bf{r}}_{j}(1\leq j \leq M)$, and 
$L_{1}$ insertions of the surface operator  
$\frac{1}{2}\vec{\phi}_{s}^{2}({\bf R}_{l})$ ($1 \leq l \leq L_{1}$). 

The corresponding parallel 
Fourier transform of the full free propagator takes the form \cite{DD81}
\be
G({{\bf{p}}},z,z') = \frac{1}{2\kappa_{0}} \left[ e^{-\kappa_{0}|z-z'|} - 
\frac{c_{0}-\kappa_{0}}{c_{0}+\kappa_{0}} e^{-\kappa_{0}(z+z')} 
\right],\label{8} 
\ee
with the standard notation 
$
\kappa_{0}=\sqrt{p^{2}+m_{0}^{2}}.
$
Here, ${\bf p}$ is the value of the parallel "momentum", i.e. the 
wave-vector, associated with $d-1$ translationally invariant directions in 
the system.

In the theory of semi-infinite systems the bulk field 
$\phi({\bf x})$ and the surface field $\phi_{s}({\bf r})$ should be 
reparameterized by different uv-finite renormalization factors 
\cite{D86,DSh98}, $Z_{\phi}(u,v)$ and $Z_{1}^{sp}(u,v)$. Thus, we have 
$ \phi = Z_{\phi}^{1/2}\phi_{R}$ and  
 $\phi_{s}=Z_{\phi}^{1/2}(Z_{1}^{sp})^{1/2}\phi_{s,R}$, where $\phi_{R}$ 
 and $\phi_{s,R}$ are the renormalized bulk and surface fields, 
 respectively. 
Besides, introduction of the additional surface operator insertions 
$\frac{1}{2}\vec{\phi}_{s}^{2}({\bf R}_{l})$ requires additional specific 
renormalization factor $Z_{\phi_{s}^2}$,  
$$\phi_{s}^{2}=[Z_{\phi_{s}^2}]^{-1}\phi^{2}_{s,R}.$$

The corresponding renormalized correlation functions involving N bulk 
fields, M surface fields and $L_{1}$ surface operators 
$\frac{1}{2}\vec{\phi}_{s}^{2}({\bf R}_{l})$ can be written as 

\be
G_{R}^{(N,M,L_{1})} (0 ; m,u,v,c)=Z_{\phi}^{-(N+M)/2} (Z_{1}^{sp})^{-M/2} 
Z_{\phi_{s}^{2}}^{L_{1}} G^{(N,M,L_{1})} (0 ; 
m_{0},u_{0},v_{0},c_{0}).\label{9} 
\ee

In the present paper we concentrate our attention on the correlation 
function $G^{(0,2,1)} (0 ; m,u,v,c)$ involving two surface fields and 
a single surface operator insertion $\vec{\phi}_{s}^{2}({\bf 
R}_{l})$.

The uv-singularities of the correlation  
function $G^{(N,M,L_{1})} (0 ; m,u,v,c)$ can be absorbed 
through a mass shift $m_{0}^{2}=m^2+\delta m^{2}$ and a surface-enhancement 
shift $c_{0}=c+\delta c$ (see \cite{DSh98}). The renormalizations of the 
mass $m$, coupling constants $u, v$ and the renormalization factor 
$Z_{\phi}$ are defined by standard normalization conditions of the 
infinite-volume theory \cite{BGZ76,P80,PS00,PV00}. In order to absorb uv 
singularities located in the vicinity of the surface, a surface-enhancement 
shift $\delta c$ is required (see Appendix 1). 
From Eq.(\ref{12}) and the expression for the renormalized correlation 
function (\ref{9}) it follows that 
\be
[Z_{\phi_{s}^2}]^{-1}=\left. Z_{\parallel}\frac{\partial 
[G^{(0,2)}(0;m_{0},u_{0},v_{0},c_{0})]^{-1}}{\partial 
c_{0}}\right|_{c_{0}=c_{0}(c,m,u,v)}.\label{14} 
\ee

It should be mentioned that the renormalization factor $Z_{\parallel} = 
Z_{1}^{sp} Z_{\phi}$ is defined via the standard normalization condition 
(\ref{11}) (see \cite{DSh98,UH02}) 
 
\be
Z_{\parallel} = \left. 2m \frac{\partial}{\partial p^{2}} 
    [G^{(0,2)} (p)]^{-1} \right|_{p^2=0} 
= \lim_{p\to 0} \frac{m}{p}
    \frac{\partial}{\partial p} [G^{(0,2)}(p)]^{-1}.
\label{15} 
\ee

It should be mentioned that all the $Z$ factors in the $d<4$ case  
have finite limits at $\Lambda \to \infty$ (where $\Lambda$ is the 
large-momentum cutoff) and depend on 
the dimensionless variables $u$ and $v$. Besides, the surface 
renormalization factors $Z_{1}^{sp}$ and $Z_{\phi_{s}^2}$ depend on both 
$u,v$ and the dimensionless ratio $c/m$. The dependence on the ratio 
$c/m$ plays a crucial role in the investigation of the crossover behavior 
from the special surface transition ($c/m\to 0$) to the ordinary 
 surface transition ($c/m\to \infty$). 

\renewcommand{\theequation}{\arabic{section}.\arabic{equation}} 
\section{Expansion of the correlation functions near the multicritical 
point} \setcounter{equation}{0}

As was indicated before, the main goal of the present work is to investigate 
the critical behavior at the special surface transition and to perform the 
analysis of the scaling critical behavior between the special and 
the ordinary transition. In this 
connection let us consider the small deviation $\Delta 
c_{0}=c_{0}-c_{sp}^{*}$ from the multicritical point. The power expansion of 
the bare correlation functions $G^{(N,M)}(0;m_{0},u_{0},v_{0},c_{0})$ in 
terms of  this small deviation $\Delta c_{0}$ has the form 
\be 
G^{(N,M)}(0;m_{0},u_{0},v_{0},c_{0})=\sum_{L_{1}=0}^{\infty} \frac{(\Delta 
c_{0})^{L_{1}}}{L_{1} !} 
G^{(N,M,L_{1})}(0;m_{0},u_{0},v_{0},c_{sp}^{*}).\label{16} \ee 
Reexpressing the right-hand side of Eq.(\ref{16}) according to the 
Eq.(\ref{9}) in terms of the renormalized correlation functions 
 and  renormalized variable 
 $ \Delta c=[Z_{\phi^{2}_{s}}(u,v)]^{-1} \Delta c_{0} $
, we obtain 
\begin{eqnarray}
&& Z_{\phi}^{-(N+M)/2} 
(Z_{1}^{sp})^{-M/2}G^{(N,M)}(0;m_{0},u_{0},v_{0},c_{0})\nonumber\\
&& = \sum_{L_{1}=0}^{\infty}\frac{(\Delta c)^{L_{1}}}{L_{1} !} 
G^{(N,M,L_{1})}_{R}(0;m,u,v).\label{18} \end{eqnarray}

The above equation in a straightforward fashion defines the corresponding  
renormalized correlation functions, defined in the vicinity of the 
multicritical point 

\be
G^{(N,M)}_{R}(0;m,u,v,\Delta c)=Z_{\phi}^{-(N+M)/2} 
(Z_{1}^{sp})^{-M/2}G^{(N,M)}(0;m_{0},u_{0},v_{0},c_{0}).\label{19}
\ee 
For dimensional reasons, we can introduce the dimensionless variable $ 
\bar{c}=\Delta c/m.$ Thus, the above correlation functions 
$G^{(N,M)}_{R}(0;m,u,v,\bar c)$ satisfy the corresponding Callan-Symanzik 
equations \cite{Sh97,DSh98} 
\begin{eqnarray}
&& \left[ m\frac{\partial}{\partial m}+\beta_{u} 
(u,v)\frac{\partial}{\partial u}+ \beta_{v} (u,v)\frac{\partial}{\partial 
v}+\frac{N+M}{2}\eta_{\phi}(u,v)\right.\nonumber\\ 
&& +\left.\frac{M}{2}\eta^{sp}_{1}(u,v)-[1+\eta_{\bar{c}}(u,v)] 
\bar{c}\frac{\partial}{\partial \bar{c}}\right] 
G^{(N,M)}_{R}(0;m,u,v,\bar c)= \Delta G_{R},\label{21}
\end{eqnarray}
where the inhomogeneous part $\Delta G_{R}$ should be negligible in the 
critical region, similarly as this took place in the case of infinite-system 
 field theory. The functions $\beta_{u}(u,v)$, 
$\beta_{v}(u,v)$ and $\eta_{\phi}(u,v)$, appearing in (\ref{21}), are the 
usual bulk RG functions \cite{Amit,P80,CPV00}.
The resulting homogeneous equation differs from the standard bulk 
Callan-Symanzik equation \cite{Z89,Parisi88,ID89} in that fashion that it 
involves the additional surface-related function $\eta^{sp}_{1}$ and the 
term $-[1+\eta_{\bar{c}}(u,v)] \bar{c}\frac{\partial}{\partial \bar{c}}$, 
where \be
\eta_{1}^{sp}(u,v)=\left. m\frac{\partial}{\partial m}\right|_{FP} ln 
Z_{1}^{sp}(u,v)=\beta_{u}(u,v)\frac{\partial ln Z_{1}^{sp}(u,v)}{\partial 
u}+ \beta_{v}(u,v)\frac{\partial ln Z_{1}^{sp}(u,v)}{\partial v}\label{22} 
\ee 
and
\be
\eta_{\bar{c}}(u,v)=\left. m\frac{\partial}{\partial m}\right|_{FP} ln 
Z_{\phi_{s}^2}(u,v)=\beta_{u}(u,v)\frac{\partial ln 
Z_{\phi_{s}^2}(u,v)}{\partial u}+ \beta_{v}(u,v)\frac{\partial ln 
Z_{\phi_{s}^2}(u,v)}{\partial v}.\label{23} \ee

In the case $\Delta c=0$ we obtain the analog of the Callan-Symanzik 
equation for the correlation functions $G^{(N,M)}_{R}(0;m,u,v)$ at the 
special point $c_{0}=c_{sp}^{*}$. 

The symbol 'FP' indicates that the above value should be calculated at the 
corresponding infrared-stable fixed point (FP) of the underlying bulk 
theory.

\renewcommand{\theequation}{\arabic{section}.\arabic{equation}} 
\section{Analysis of the scaling critical behavior} 
\setcounter{equation}{0}

The asymptotic scaling critical behavior of the correlation functions 
can be obtained through detailed analysis of the CS equation 
(\ref{21}), as was proposed in \cite{Z89,BB81} and 
employed to the case of the semi-infinite systems in 
\cite{DSh98,CR97,MCD99}. In the critical region, at 
$\tau=(T-T_{c})/T_{c}\to 0$, for the renormalization Z factors we obtain
\begin{eqnarray}
&& Z_{\phi}\sim m^{\eta_{\phi}(u^{*},v^{*})},\nonumber\\
&& Z_{1}^{sp}\sim m^{\eta^{sp}_{1}(u^{*},v^{*})},\nonumber\\
&& Z_{\phi_{s}^{2}}\sim 
m^{\eta_{\bar{c}}(u^{*},v^{*})},\label{25}
\end{eqnarray}
where the variable $m$ is identified with 
the inverse bulk correlation length $\xi^{-1}\sim 
\tau^{\nu}$, as it is usually accepted in the massive field theory 
\cite{P80,Z89,Parisi88}.

Substituting the last equation from (\ref{25}) into the expressions for 
$\Delta c$ and  for the scaling variable $\bar{c}$ it is easy to obtain  the 
following asymptotic forms
\be
\Delta c\sim m^{-\eta_{\bar{c}}(u^{*},v^{*})} \Delta 
c_{0},\quad\quad \Delta c\sim \tau^{-\nu 
\eta_{\bar{c}}(u^{*},v^{*})} \Delta c_{0}\label{26} \ee
and
\be
\bar{c}\sim m^{-(1+\eta_{\bar{c}}(u^{*},v^{*}))} \Delta 
c_{0},\quad\quad\quad \bar{c}\sim \tau^{-\Phi} \Delta c_{0},\label{27}
\ee
where
\be
\Phi=\nu 
(1+\eta_{\bar{c}}(u^{*},v^{*})) \label{28}
\ee
is the surface crossover critical exponent. The second equation in 
(\ref{27}) explains the physical meaning of the surface crossover exponent 
as a value characterising the measure of deviation from the multicritical 
point.  The second equations in (\ref{26}) and (\ref{27}) indicate 
non-analitic temperature dependence of the renormalized surface-enhancement 
deviation $\Delta c$. Thus, from the analysis of the CS equation (\ref{21}) 
we obtain the following asymptotic scaling form of the surface correlation 
function $G^{(0,2)}$ :

\begin{eqnarray}
&& G^{(0,2)}(p;m_{0},u_{0},v_{0},c_{0})\sim 
m^{-\frac{\gamma_{11}^{sp}}{\nu}} 
G^{(0,2)}_{R}(\frac{p}{m};1,u^{*},v^{*},m^{-\Phi / \nu}\Delta 
c_{0})\nonumber\\ && \sim 
\tau^{-\gamma_{11}^{sp}}G(pt^{-\nu};1,\tau^{-\Phi}\Delta c_{0}), \label{29} 
\end{eqnarray} where
$
\gamma_{11}^{sp}=\nu (1-\eta_{\parallel}), 
$
is the local surface susceptebility exponent and 
\be
\eta_{\parallel}^{sp}=\eta_{1}^{sp}+\eta_{\phi}\label{30}
\ee
is the surface correlation exponent at the special surface transition.
It is easy to see that the asymptotic scaling critical 
behavior of the surface correlation function for the semi-infinite 
$n$-component systems with cubic anisotropy is characterized by the new 
crossover exponent $\Phi(u^{*},v^{*})$, calculated at the cubic fixed 
point $(u^{*},v^{*})$.

\renewcommand{\theequation}{\arabic{section}.\arabic{equation}}
\section{The fixed-dimension perturbative expansion up to two-loops.} 
\setcounter{equation}{0}

As was investigated previously \cite{DSh94,DSh98,UShH00,UH02}, the 
fixed-dimension $\phi^4$ field-theoretic approach \cite{P80} provides 
an accurate description of the surface critical behavior of semi-infinite 
systems. We apply this method to the analysis of the cubic anisotropic model 
(\ref{2}) at the special surface transition and to the invetigation of the 
crossover behavior in the vicinity of the multicritical point 
$c_{0}=c_{sp}^{*}$. The surface correlation exponent 
$\eta_{\parallel}^{sp}$ at the special transition can be obtained from  
Eq.(\ref{30}), where $\eta_{1}^{sp}$ is defined by Eq.(\ref{22}), and 
$\eta_{\phi}(u,v)$ is the standart bulk exponent $\eta_{\phi}=\left. 
m\frac{\partial}{\partial m}ln Z_{\phi}\right|_{FP}$. After performance 
of the mass- and surface-enhancement renormalization and carrying out the 
integration of the corresponding Feynman integrals with subsequent 
execution of the standart vertex renormalizations of bare dimensionless 
parameters 
$\bar{u}_{0}=\bar{u}(1+\frac{n+8}{6}\bar{u}+\bar{v})$,
$\bar{v}_{0}=\bar{v}(1+\frac{3}{2}\bar{v}+2\bar{u})$
(where $\bar{u}_{0}=u_{0}/8\pi m$ and $\bar{v}_{0}=v_{0}/8\pi m$),
by analogy to that as it was done in \cite{UH02} for random semi-infinite 
model, we obtain in the case of the cubic anisotropic model the following 
expression for the surface correlation exponent at the
special transition 
\begin{eqnarray}
\label{etafin}
&&\eta_{\parallel}^{sp}(u,v)=
	-\frac{n{+}2}{2(n{+}8)} \, u - \frac{v}{6}\\ 
&& + 12 \frac{(n{+}2)}{(n{+}8)^2} A(n) u^2 + \frac{4}{9} 
    A(1) v^2 + \frac{8}{n{+}8} A(n) uv. \nonumber
\end{eqnarray}
Here $A(n)$ is defined as 
\begin{equation}
A(n)=2A+{n{-}10\over 48}+\frac{n+2}{6}(ln^{2} 2 - ln 2),
\end{equation}
and the renormalized coupling constants $u$ and $v$, are normalized in the  
standard fashion, $u{=}{n+8\over 6}{\bar{u}}$ and $v{=}{3\over 
2}{\bar{v}}$. 

According to the Eq.(\ref{28}) and Eq.(\ref{23}), the calculation 
of the crossover critical exponent $\Phi$ is connected with the calculation 
of the renormalization factor $Z_{\phi_{s}^2}$ via (\ref{14}). 
With that end in view we can 
rewrite the normalization condition (\ref{10}) in the form 
\be
Z_{\parallel}[G^{(0,2)}(0;m_0,u_0,v_0,c_0)]^{-1}=m + c\label{34}
\ee
for the inverse unrenormalized surface correlation function 
$[G^{(0,2)}(0)]^{-1}$. It is easy to see that by differentiation 
of the above mentioned normalization condition with respect to $c_{0}$ and 
by taking into account Eq.(\ref{14}) we obtain for the renormalization 
factor $Z_{\phi^2}$ the equation \be
Z_{\phi_{s}^2}=\frac{\partial c_{0}}{\partial c},\label{35}
\ee
where $c_{0}=c+\delta c$ and 
\be
\delta c = (Z_{\parallel}^{-1}-1)(m+c) + \sigma_{0}(0;m,c_0=c+\delta 
c).
\label{36}
\ee
Here $\sigma_{0}(0;m,c_0)=\sigma_{1}+\sigma_{2}+\sigma_{3}+\sigma_{4}$ 
denotes the sum of loop diagrams of all orders in 
$[G^{(0,2)}(0;m,u_{0},v_{0},c_{0})]^{-1}$ (see \cite{DSh98,UH02}). Among 
them $\sigma_{1}$ corresponds to the one-loop graph, $\sigma_{2}$ denotes 
the melon-like two-loop diagrams \be
\sigma_{2}= \raisebox{-5pt}{\epsfxsize=1.7cm \epsfbox{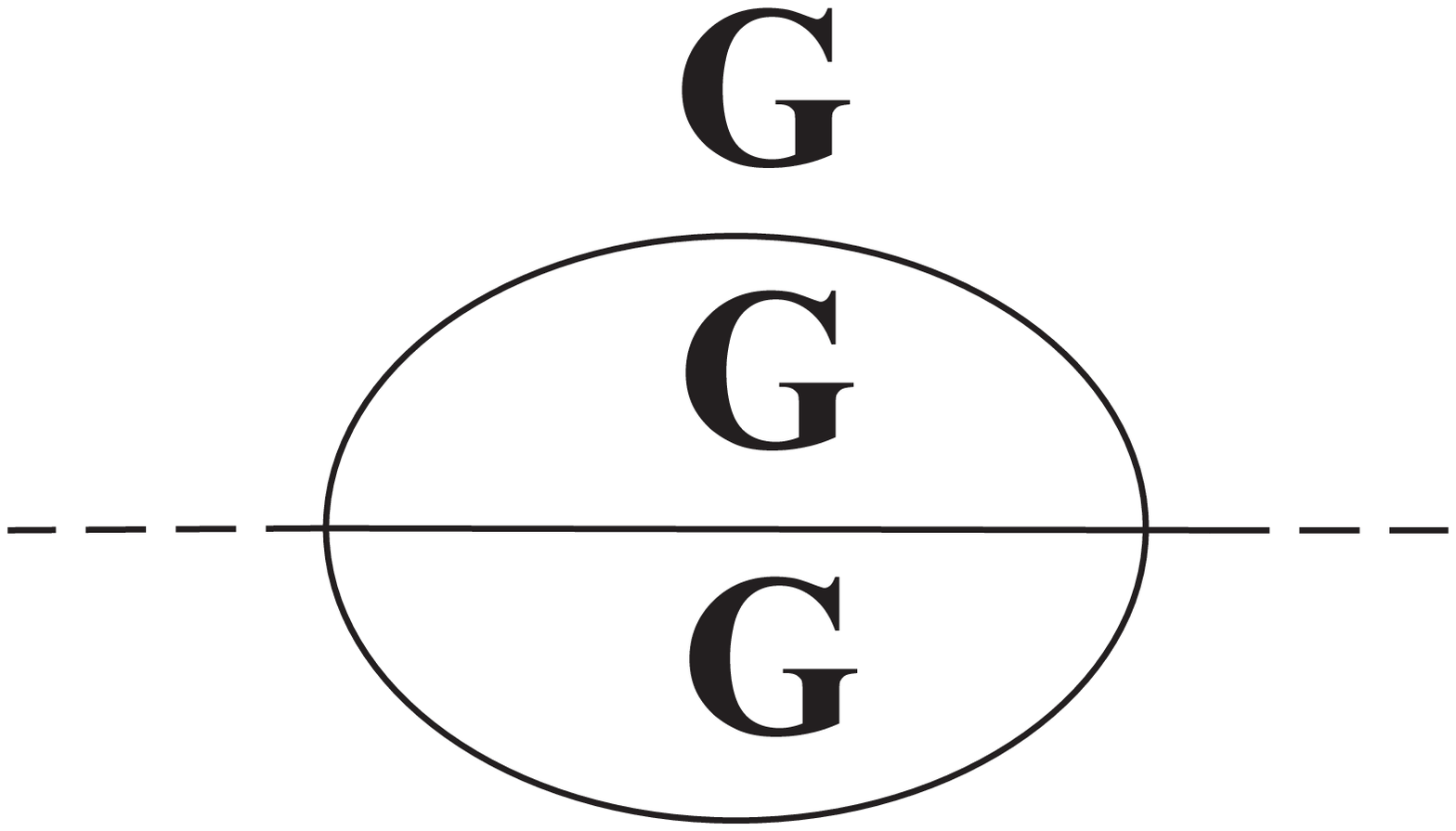}}
-\frac{1}{2\kappa}\raisebox{-5pt}{\epsfxsize=0.9cm \epsfbox{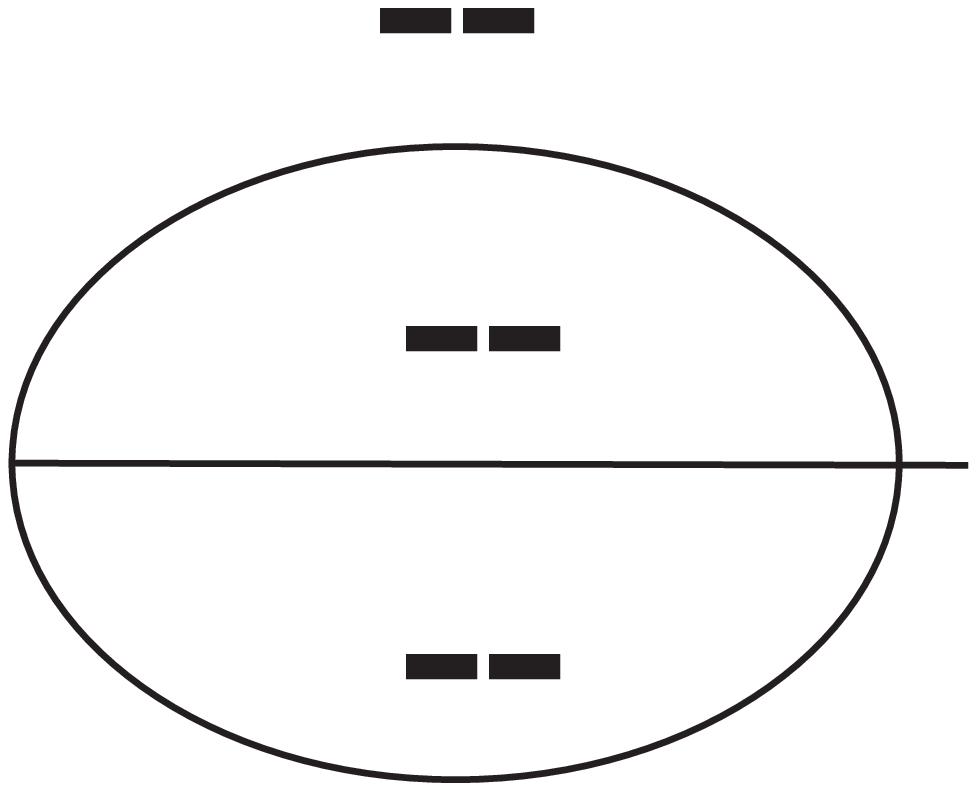}} +
\frac{m^2}{2\kappa}\frac{\partial}{\partial k^2}\left. 
\raisebox{-5pt}{\epsfxsize=1cm 
\epsfbox{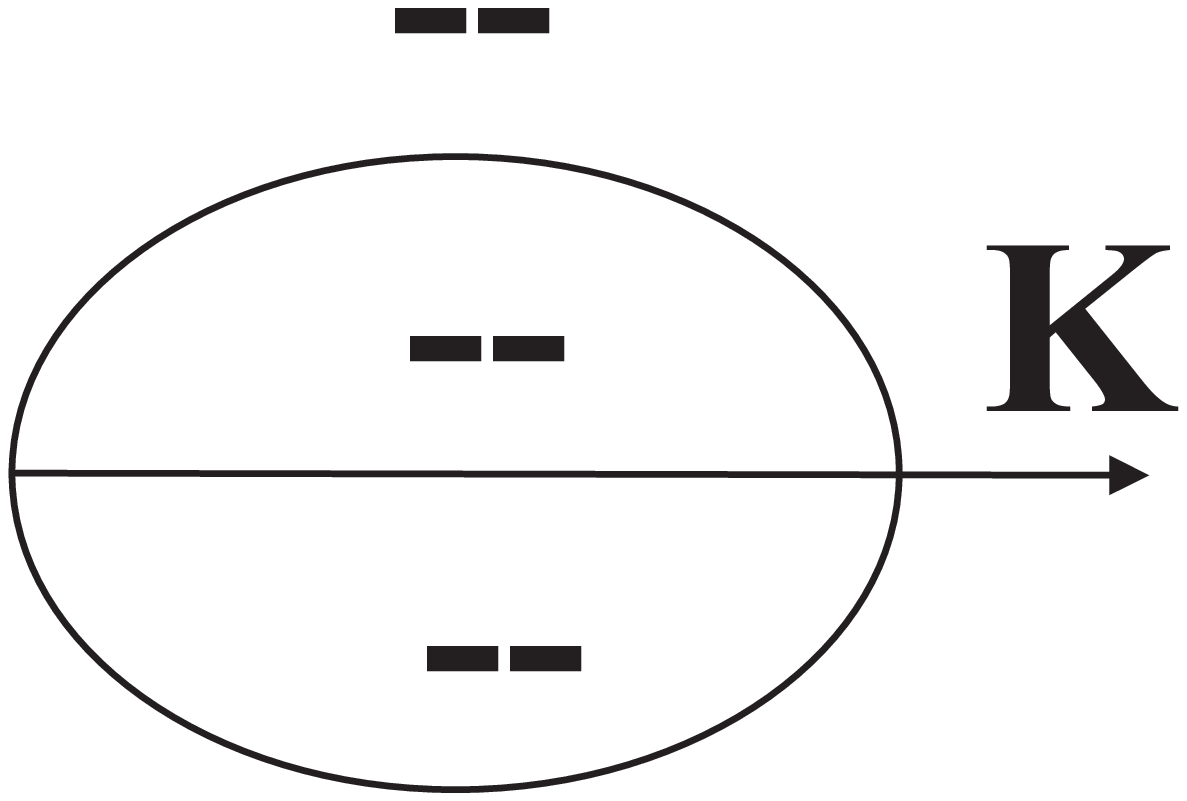}}\right|_{k^2=0},\label{37}
\ee

$\sigma_{3}$ and $\sigma_{4}$ represent the one-particle reducible and 
one-particle irreducible two-loop diagrams in 
$[G^{(0,2)}(0;m,u_{0},v_{0},c_{0})]^{-1}$, respectively. These graphs have 
their correponding weights (arise from the standard symmetry factors)
\begin{subequations}
\label{ti}
\begin{eqnarray}
&-&{t_1^{(0)}\over 2}\quad\mbox{with}\quad t_1^{(0)}=\frac{n+2}{3}\,u_0+ v_{0}\,,\\
&&{t_2^{(0)}\over 6}\quad\mbox{with}\quad 
t_2^{(0)}=\frac{n+2}{3}\, u_{0}^{2}+ v_{0}^{2}+ 2v_{0}u_{0}\,,\\
&&{t_3^{(0)}\over 4}\quad\mbox{and}\quad {t_4^{(0)}\over 
4}\quad\mbox{with}\quad t_3^{(0)}=t_4^{(0)}=\big(t_1^{(0)}\big)^2.
\end{eqnarray}
\end{subequations}

The full lines 
with notes "G" denote the full free propagator (\ref{8}) and signs "-" 
denote the free bulk propagators, which are associated with the first 
term in the full free propagator (\ref{8}). The equation 
(\ref{35}) can be resolved by using the method of sequential iteration. As 
the result, at the second order of the perturbation theory, we obtain 
the renormalization factor $Z_{\phi_{s}^2}$ in terms of the new renormalized 
coupling constants $\bar{u}$ and $\bar{v}$, 
\be
Z_{\phi_{s}^2}(\bar{u},\bar{v})=1+\frac{n+2}{3}(ln2-\frac{1}{4})\bar{u}+
(ln2-\frac{1}{4})\bar{v}+\frac{n+2}{3}C(n)\bar{u}^2+2C(n)\bar{u}\bar{v}+C(1)\bar{v}^2,\label{44}
\ee
where $C(n)$ is a function defined as 
\be
C(n)=A-B-\frac{n}{2}ln2+\frac{n+2}{2}ln^{2}2+\frac{2n+1}{12},\label{45}
\ee

and $A=0.202428$, $B=0.678061$ are 
integrals originating from the two-loop melon-like diagrams. 
Combining the renormalization 
factor $Z_{\phi_{s}^2}$ with the one-loop pieces of the $\beta$ functions 
$\beta_{\bar{u}}(\bar{u},\bar{v})=-\bar{u}(1-[(n+8)/6]\bar{u}-\bar{v})$ and 
$\beta_{\bar{v}}(\bar{u},\bar{v})=-\bar{v}(1-\frac{3}{2}\bar{v}-2\bar{u}),$ 
according to Eq.(\ref{23}), we obtain the desired series expansion for 
$\eta_{\bar{c}}$, 
\be
\eta_{\bar{c}}(u,v)=-2\frac{n+2}{n+8}(ln2-\frac{1}{4})u-\frac{2}{3}(ln2-\frac{1}{4})v-
8[3\frac{n+2}{(n+8)^2}D(n)u^2+\frac{2D(n)}{n+8}uv+\frac{D(1)}{9}v^2],\label{46}
\ee 
where 
\be
D(n)=A-B+\frac{n+2}{3}ln^{2}2-\frac{n+1}{2}ln2+\frac{17n+22}{96}.\label{47}
\ee
Eq.(\ref{46}) gives a result for the model with the effective Hamiltonian 
of the Landau-Ginzburg-Wilson type with cubic anisotropy in the 
semi-infinite space (\ref{2}), with a general number $n$ of order parameter 
components. 

The knowledge of $\eta_{\bar{c}}$ gives an access to the calculation of 
the crossover critical exponent $\Phi$ via the scaling relation (\ref{28}). 
Additionally, we can calculate the critical exponents $\alpha_{1}$ and 
$\alpha_{11}$ of the layer and specific heats via the usual scaling 
relations \cite{D86}
\be
\alpha_{1}=\alpha+\nu-1+\Phi=1-\nu (d-2-\eta_{\bar{c}}),\quad\quad 
\alpha_{11}=\alpha+\nu-2+2\Phi = -\nu 
[d-3-2\eta_{\bar{c}}].\label{49} 
\ee

The above critical exponents should be calculated for different $n$ 
($n=3,4,8$, and $n \to\infty$) at the standard infrared-stable cubic fixed 
(FP) points of the underlying bulk theory, as it is usually accepted in the 
massive field theory. As it was mentioned above, in the cases $n<n_{c}$ 
the cubic ferromagnets are described by the Heisenberg isotropic Hamiltonian 
at the $O(n)$-symmetric fixed point.

In the case of the replica limit $n\to 0$ we can obtain from (\ref{46}) the 
series of $\eta_{\bar{c}}^{r}$ and corresponding expressions for the surface 
crossover critical exponent $\Phi^{r}$, and critical exponents 
$\alpha_{1}^{r}$ and $\alpha_{11}^{r}$ of the layer and specific heats for 
{\it{semi-infinite random Ising-like}} systems (\cite{UH03})

\section{Numerical results}

In order to obtain the full set of surface critical exponents at the 
special transition and to calculate the surface crossover exponent $\Phi$ 
from the special to the ordinary transition in systems with cubic anisotropy 
we substitute the expansion (\ref{etafin}) for $\eta_{\parallel}^{sp}$ into 
the standart scaling-law expressions for the surface exponents (see Appendix 
2) and the expansion (\ref{46}) for $\eta_{\bar{c}}$ into the scaling 
relations (\ref{28}) and (\ref{49}). 

For each of the above mentioned surface critical exponents of the special 
transition and for the crossover exponent $\Phi$ we obtain at $d=3$ a double 
series expansion in powers of $u$ and $v$, truncated at the second order.
As it is known \cite{mit3,B87,Mc94,AMR00}, power series expansions of this 
kind are generally divergent due to a nearly factorial growth of expansion 
coefficients at large orders of perturbation theory.
In order to perform the analysis of these perturbative series expansions and 
to obtain accurate estimates of the surface critical exponents, a powerful 
resummation procedure must be used. One of the simplest ways is to perform 
the double Pad\'e-analysis \cite{B75}. This 
should work well when the series behave in lowest orders "in a convergent 
fashion". Another way is to perform for these series the double Pad\'e-Borel 
analysis \cite{BNGM76}. The Pad\'e-Borel 
resummation procedures are posible to use only in the case when the series 
are alternating in sign \cite{BNM78}. 
 
The results of our calculations of the 
surface critical exponents of the special transition for various values of 
$n=3,4,8,\infty$ at corresponding cubic fixed points are presented in 
Tables 1-5. 
Unfortunately, the second ($p=2$) order analysis of perturbative series 
\cite{mitSh88} gives the cubic fixed point with coordinates $u_{0}=1.5347$ 
and $v_{0}=-0.0674$ at $n=3$ for the 3D model.
 This corresponds to ordering along the easy axes, because 
$(v_{0}<0)$. The analysis of the eigenvalues of the stability matrix shows 
that in the frames of the two-loop approximation the cubic fixed point at 
$n=3$ is unstable and the $O(n)$-symetric fixed point is stable. 
But, the estimations of 
 the marginal spin dimensionality of the cubic model $n_{c}$ in the 
 frames of three-loop \cite{MayerSokolov87,Sh89}, 
 four-loop \cite{MSSh89,V00}, five-loop $\epsilon=4-d$ expansion 
\cite{KSch95,ShAS97,V00} and six-loop study  
\cite{CPV00} show that the cubic ferromagnets are not described by the 
Heisenberg isotropic Hamiltonian, but by the cubic model at the stable cubic 
fixed point. Higher precision six-loop field-theoretical analysis 
\cite{CPV00} give the value of the marginal spin dimensionality of the cubic 
model equal to $n_{c}=2.89(4)$.  In accordance with 
this we used the cubic fixed point of the higher $p=3$ order of perturbative 
series for obtaining the set of surface critical exponents at $n=3$. For 
estimation of the reliability of the obtained results we performed 
calculations at the cubic fixed point of the $p=6$ order in Table 2. The 
obtained results indicate that the difference in the ways of the $\beta$ 
functions resummation have no essential influence on the values of the 
surface critical exponents and that the results obtained in the frames of 
the two-loop approximation are stable and reliable. The surface critical 
exponents of the special transition for $n=4,8$ and $n\to\infty$ were 
calculated at the standard infrared-stable cubic fixed (FP) points of the 
underlying bulk theory, as it is usually accepted in the massive field 
theory. 

The Tables 6-9 represent the surface crossover exponent $\Phi$ 
and the surface critical exponents of the layer $\alpha_{1}$ and local 
$\alpha_{11}$ specific heats for different values of $n=3,4,8,\infty$ at 
the corresponding cubic fixed points.

 The quantities $O_{1}/O_{2}$ and $O_{1i}/O_{2i}$ represent the ratios of 
magnitudes of first-order and second-order perturbative corrections 
appearing in direct and inverse series expansions. The larger (absolute) 
value of these ratios indicate the better apparent convergence of 
truncated series.
 
The values $[p/q]$ (where $p,q=0,1$) are 
simply Pad\'e approximants which represent the partial sums of the direct 
and inverse series expansions up to the first and the second order. The 
nearly diagonal two-variable rational approximants of the types $[11/1]$ and 
$[1/11]$ give at $u=0$ or $v=0$ the usual 
$[1/1]$ Pad\'e approximant \cite{B75}. The results of Pad\'e-Borel analysis 
of the direct $R$ and the inverse $R^{-1}$ series expansions give numerical 
estimates of the surface critical exponents with a high degree of 
reliability. As it is easy to see from Tables 1-5, the most reliable 
estimate is obtained from the inverse series expansion for the surface 
critical exponent $\eta_{\perp}$, which represent the best convergence 
properties. Thus we obtain for $\eta_{\perp}$:
$-0.121$ at $n=3$;
$-0.139$ at $n=4$;
$-0.143$ at $n=8$;  
$-0.141$ at $n\to \infty$.
Substituting these values of $\eta_{\perp}$ together with the standard bulk 
values $\nu$ ($\nu=0.697$ \cite{Sh89} (or $0.706(6)$ at $p=6$ \cite{CPV00}) 
$(n=3)$; $0.711$ $(n=4)$; $0.717$ $(n=8)$; $0.715$ $(n\to\infty)$) 
\cite{mit2} and $\eta$ ($\eta=0.024$ \cite{Sh89} (or $0.0333(26)$ at $p=6$ 
\cite{CPV00}) $(n=3)$; $0.026$ $(n=4)$ ; $0.025$ $(n=8)$; $0.025$ 
$(n\to\infty)$)\cite{mit2} at $d=3$ into the surface scaling relations 
(see Appendix 2) we have obtained the remaining critical exponents that are 
present in the last columns of Tables 1-5. The deviations of these estimates 
from the other estimates of the table give a rough measure of the achieved 
numerical accuracy. 

The double Pad\'e and Pad\'e-Borel analysis of the series 
for $\eta_{\bar{c}}$, $\alpha_{1}$, $\alpha_{11}$ and $\Phi$, presented in 
Tables 6-9, was performed in a similar way. In this case the situation is more 
complicated, because the series of these critical exponents exhibit bad 
convergence properties. A very similar situation took place in the 
analysis of perturbation expansions for the surface critical exponents in 
isotropic systems \cite{DSh98}. As it is easy to see, we obtain the most 
reliable estimates for the surface critical exponent $\alpha_{11}$. The 
result of substituting of this value $\alpha_{11}$ ($\alpha_{11}=-0.331$ 
$(n=3)$; $-0.374$ $(n=4)$; $-0.384$ $(n=8)$; $-0.380$ $(n\to\infty)$), 
together with the value of $\nu$, into scaling laws (\ref{28}) and 
(\ref{49}) is presented in the last columns of Tables 6-9. For evaluation 
of the reliability of the obtained results we have performed additional 
calculation of the surface critical exponents on the basis of these 
$\alpha_{11}$ and the corresponding six-loop perturbation theory results 
\cite{CPV00} for bulk critical exponent of the correlation length 
$\nu=0.706(6)$ $(n=3)$; $0.714(8)$ $(n=4)$; $0.712(6)$ $(n=8)$; $0.708(8)$ 
$(n\to\infty)$. The results of calculation are presented in Table 10. As it 
is easy to see, these results differ very little from the 
 results presented in the last columns of Tables 6-9. It indicates 
good stability of the results obtained in the frames of the two-loop 
approximation scheme. 
 
The obtained results for surface critical exponents of 
semi-infinite model with cubic anisotropy, calculated at the cubic fixed 
point are different from the results for surface critical exponents of 
standart semi-infinite $n$-component model (see 
\cite{DD81,DD83,DSh94,DSh98}). For example, the 
difference in the case of $d=3$ at $n=3$ is about $7.5 \%$ for 
$\eta_{\parallel}$ and $6.6 \%$ for $\eta_{\perp}$ \cite{mit5}. 

If $n<n_{c}$, the cubic fixed point is unstable and 
the cubic term in the Hamiltonian (\ref{2}) becomes irrelevant. In this case 
the isotropic fixed point is stable and the system is described by the 
simple $O(n)$-symmetric model in 3D. The corresponding  
 surface critical exponents can be 
calculated from the series, presented in \cite{DSh98}.

As was indicated previously, in the limit $n\to 0$, the cubic model 
(\ref{2}) with $u_{0}<0$ and $v_{0}>0$ describes the semi-infinite 
Ising-like systems with random bulk disorder. The investigation of the 
surface special transition for such kind of systems was presented in 
\cite{UH02}. For the value of $\eta_{\bar{c}}^{r}$, for the surface 
crossover exponent $\Phi^{r}$ and for the surface exponents of the layer 
$\alpha_{1}^{r}$ and local $\alpha_{11}^{r}$ specific heat, in the 
limit $n\to 0$ at the standard infrared-stable random fixed point 
$u^{*}=-0.60509$ $v^{*}=2.39631$ \cite{Jug83} of the underlying bulk theory 
we obtain \cite{UH03} 
\begin{eqnarray}
&& \eta_{\bar{c}}^{r}=-0.164,\quad\quad\quad
\alpha_{1}^{r}=0.211,\nonumber\\
&&\alpha_{11}^{r}=-0.222\quad\quad\quad \Phi^{r}=0.567.
\end{eqnarray} 
These values are different from the surface critical exponents of the pure 
semi-infinite Ising-like systems and indicate that in a system with random 
bulk disorder the planar boundary is characterized by a new set of the 
surface critical exponents.

\section{Summary} 

We have studied special surface transition and crosover critical behavior 
in the vicinity of the multicritical point for 3D semi-infinite systems with 
cubic anisotropy by applying the 
field theoretic approach directly in $d=3$ dimensions up to the two-loop 
approximation. We have performed a double Pad\'e and Pad\'e-Borel analysis 
of the resulting perturbation series 
for the surface critical exponents of the special 
transition, the surface crossover exponent $\Phi$ and the critical exponents 
of the layer $\alpha_{1}$ and local specific heats $\alpha_{11}$ for various 
$n=3,4,8,\infty$, in order to find the best numerical estimates.

We find that at $n>n_{c}$ the surface critical exponents of the special 
transition in semi-infinite systems with cubic anisotropy belong to the 
cubic universality class and the asymptotic scaling critical behavior of 
the surface correlation function in the vicinity of the multicritical 
point is characterized by the new crossover exponent of 
the cubic universality class $\Phi(u^{*},v^{*})$.

Besides, we present the results of investigation of the crossover 
critical behavior between special and ordinary transition in random 
semi-infinite systems, by taking the limit $n\to 0$ \cite{mit4}. In this 
case we calculate the surface critical exponents $\alpha_{1}^{r}$, 
$\alpha_{11}^{r}$ and the surface crossover exponent $\Phi^{r}$ and 
confirm that they belong to the universality class of the random model.

The further theoretical investigation of the asymptotic surface critical 
behavior of semi-infinite cubic systems will be highly desirable in the 
framework of higher-order (three-loop) RG approximations. 
In particular, it would be very interesting to study the case of the 3D 
cubic crystal with $n=3$.

We suggest that the obtained results could stimulate further experimental 
and numerical investigations of the surface critical behavior of random 
systems and systems with cubic anisotropy.

\section*{Acknowledgments}

I would like to thank Dr.A.Ciach for a useful discussion and for reading the 
manuscript. I also would like to thank Dr.A.Ciach for her hospitality at the 
Institute of Physical Chemistry of the Polish Academy of Sciences. 
 This work was supported in part by NATO Science Fellowships National 
Administration under Grant No. 14/B/02/PL. 

\renewcommand{\theequation}{A1.\arabic{equation}}
\section*{Appendix 1}
\setcounter{equation}{0}

We require that the following surface normalization conditions 
take place \cite{DSh94,DSh98} : 
\be
G_{R}^{(0,2)}(0;m,u,v,c) = \frac{1}{m+c},\label{10}
\ee
\be
\left.\frac{\partial G_{R}^{(0,2)} (p;m,u,v,c)}{\partial p^{2}} 
\right|_{p=0} = - 
\frac{1}{2m(m+c)^{2}}, \label{11}
\ee
\be
G_{R}^{(0,2,1)}(0;m,u,v,c)=\frac{1}{(m+c)^2}\label{12}
\ee
where the correlation function 
$G^{(0,2,1)}$ contains the insertion of the surface operator 
$\frac{1}{2}\phi_{s}^2$.

The Eq.(\ref{12}) is motivated by the fact that the bare 
correlation function $G^{(0,2,1)}(0;m_{0}^2,u_{0},v_{0},c_{0})$ may be 
written as a derivative 
$-\frac{\partial}{\partial c_{0}}G^{(0,2)}(0;m_{0}^2,u_{0},v_{0},c_{0})$.
This equation simplifies considerably the 
calculation of the correlation function with 
insertions of the surface operator $\frac{1}{2}\phi_{s}^2$. 

It is easy to see from (\ref{10}) that the special point is 
located at $m=c=0$, because at this point the divergence of the bulk and the 
surface correlation length and susceptibility is observed. At $c=0$ the 
surface normalization conditions are simplified and yield 
$c_{0}=c_{sp}^{*}$. This point corresponds to the multicritical point 
$(m_{0c}^{2},c_{sp}^{*})$ at which special transition takes place. On the 
other hand, the above mentioned equation implies also that the surface 
correlation length and the susceptibility are finite at the ordinary 
transition, because in this case we have $c>0$ when $m \to 0$. This latter  
case corresponds to the situation when the surface remains "noncritical" at 
the bulk transition temperature.

\renewcommand{\theequation}{A1.\arabic{equation}}
\section*{Appendix 2}
\setcounter{equation}{0}

The individual RG series expansions for other critical exponents can be 
derived through standard surface scaling relations \cite{D86} with $d=3$ 
\begin{eqnarray} 
&& \eta_{\perp} = \frac{\eta + 
\eta_{\parallel}}{2}, \nonumber\\ 
&& \beta_{1} = \frac{\nu}{2} 
(d-2+\eta_{\parallel}), \nonumber\\ 
&& \gamma_{11}=\nu(1-\eta_{\parallel}), \nonumber\\ 
&& \gamma_{1}= \nu(2-\eta_{\perp}), \label{sc}\\
&& \Delta_{1}= \frac{\nu}{2} (d-\eta_{\parallel}), \nonumber\\ 
&& \delta_{1} = \frac{\Delta}{\beta_{1}} = 
\frac{d+2-\eta}{d-2+\eta_{\parallel}}, \nonumber\\ 
&& \delta_{11} = \frac{\Delta_{1}}{\beta_{1}}= 
\frac{d-\eta_{\parallel}}{d-2+\eta_{\parallel}}\;.\nonumber
\end{eqnarray}

Each of these critical exponents characterizes certain properties of 
the cubic anisotropic system near the external surface, in the vicinity of 
the critical point. The values $\nu$, $\eta$, and 
$\Delta=\nu(d+2-\eta)/2$ are the standard bulk exponents.

\newpage
\begin{table}[htb]
\caption{Surface critical exponents of the special transition for $d=3$ 
up to two-loop order at the cubic fixed point (of order $p=3$) 
$u^{*}=1.348, v^{*}=0.074$ at $n=3$.}
\label{tab1}
\begin{center}
\begin{tabular}{rrrrrrrrrrrrr}
\hline
$ exp $~&~$\frac{O_{1}}{O_{2}}$~&~$\frac{O_{1i}}{O_{2i}}$~&~$[0/0]$~&~
$[1/0]$~&~$[0/1]$~&~$[2/0]$~&~$[0/2]$~&~$[11/1]$~&~$[1/11]$~&~$ R $~&~
$ R^{-1} $~&~$ f(\eta_{\perp},\nu,\eta) $ 
\\ \hline
$\eta_{\parallel}$ & -4.0 & 14.7 & 0.00 & -0.319 & -0.242 & -0.239 & 
-0.254 & -0.255 & -0.255 & -0.260 & --- & -0.266 \\

$\Delta_{1}$ & 10.8 & -4.4 & 0.75 & 1.069 & 1.218 & 1.098 & 1.078 & 1.101
 & 1.102 & --- & 1.109 & 1.138 \\

$\eta_{\perp}$ & -3.1 & -6.0 & 0.00 & -0.159 & -0.137 & -0.107 & -0.117 & 
-0.121 & -0.120 & -0.124 & -0.121 & -0.121 \\

$\beta_{1}$ & 0.0 & 0.0 & 0.25 & 0.25 & 0.25 & 0.253 & 0.253 & 
--- & --- & --- & --- & 0.256 \\

$\gamma_{11}$ & 11.8 & -4.3 & 0.50 & 0.819 & 0.968 & 0.846 & 0.823 
& 0.848 & 0.849 & --- & 0.857 & 0.882 \\

$\gamma_{1}$ & 12.6 & -3.1 & 1.00 & 1.398 & 1.662 & 1.430 & 
1.372 & 1.433 & 1.433 & --- & 1.450 & 1.478 \\

$\delta_{1}$ & -3.8 & -1.7 & 5.00 & 6.593 & 7.339 & 6.236 & 5.763 & 
6.262 & 6.266 & 6.287 & 6.359 & 6.779 \\

$\delta_{11}$ & -5.8 & -1.7 & 3.00 & 4.275 & 5.217 & 4.057 & 3.622 & 
4.090 & 4.095 & 4.101 & 4.186 & 4.450 \\
    
\end{tabular}
\end{center}
\end{table}

\begin{table}[htb]
\caption{Surface critical exponents of the special transition for $d=3$ 
up to two-loop order at the cubic fixed point (of order $p=6$) 
$u^{*}=1.321(18), v^{*}=0.096(20)$ at $n=3$.}
\label{tab2}
\begin{center}
\begin{tabular}{rrrrrrrrrrrrr}
\hline
$ exp $~&~$\frac{O_{1}}{O_{2}}$~&~$\frac{O_{1i}}{O_{2i}}$~&~$[0/0]$~&~
$[1/0]$~&~$[0/1]$~&~$[2/0]$~&~$[0/2]$~&~$[11/1]$~&~$[1/11]$~&~$ R $~&~
$ R^{-1} $~&~$ f(\eta_{\perp},\nu,\eta) $ 
\\ \hline
$\eta_{\parallel}$ & -4.0 & 14.9 & 0.00 & -0.316 & -0.240 & -0.237 & 
-0.252 & -0.254 & -0.253 & -0.258 & --- & -0.275 \\

$\Delta_{1}$ & 10.9 & -4.5 & 0.75 & 1.066 & 1.212 & 1.095 & 1.075 & 1.098
 & 1.099 & --- & 1.106 & 1.156 \\

$\eta_{\perp}$ & -3.1 & -6.0 & 0.00 & -0.158 & -0.137 & -0.107 & -0.117 & 
-0.120 & -0.120 & -0.123 & -0.121 & -0.121 \\

$\beta_{1}$ & 0.0 & 0.0 & 0.25 & 0.25 & 0.25 & 0.253 & 0.253 & 
--- & --- & --- & --- & 0.256 \\

$\gamma_{11}$ & 11.9 & -4.3 & 0.50 & 0.816 & 0.962 & 0.843 & 0.821 
& 0.845 & 0.846 & --- & 0.854 & 0.900 \\

$\gamma_{1}$ & 12.7 & -3.2 & 1.00 & 1.395 & 1.654 & 1.426 & 
1.370 & 1.429 & 1.430 & --- & 1.446 & 1.497 \\

$\delta_{1}$ & -3.8 & -1.7 & 5.00 & 6.581 & 7.312 & 6.248 & 5.765 & 
6.255 & 6.260 & 6.279 & 6.351 & 6.853 \\

$\delta_{11}$ & -5.9 & -1.7 & 3.00 & 4.265 & 5.187 & 4.050 & 3.624 & 
4.083 & 4.089 & 4.094 & 4.178 & 4.520 \\
    
\end{tabular}
\end{center}
\end{table}

\begin{table}[htb]
\caption{Surface critical exponents of the special transition for $d=3$ 
up to two-loop order at the cubic fixed point (of order $p=2$) $
u^{*}=1.064, v^{*}=0.520$ at $n=4$.}
\label{tab3}
\begin{center}
\begin{tabular}{rrrrrrrrrrrrr}
\hline
$ exp $~&~$\frac{O_{1}}{O_{2}}$~&~$\frac{O_{1i}}{O_{2i}}$~&~$[0/0]$~&~
$[1/0]$~&~$[0/1]$~&~$[2/0]$~&~$[0/2]$~&~$[11/1]$~&~$[1/11]$~&~$ R $~&~
$ R^{-1} $~&~$ f(\eta_{\perp},\nu,\eta) $ 
\\ \hline
$\eta_{\parallel}$ & -4.6 & 7.3 & 0.00 & -0.353 & -0.261 & -0.277 & 
-0.286 & -0.294 & -0.288 & -0.299 & --- & -0.304 \\

$\Delta_{1}$ & 7.2 & -4.7 & 0.75 & 1.103 & 1.295 & 1.152 & 1.134 & 1.157
 & 1.165 & --- & 1.173 & 1.175 \\

$\eta_{\perp}$ & -3.5 & -8.8 & 0.00 & -0.176 & -0.150 & -0.125 & -0.135 & 
-0.139 & -0.138 & -0.143 & -0.139 & -0.139 \\

$\beta_{1}$ & 0.0 & 0.0 & 0.25 & 0.25 & 0.25 & 0.254 & 0.254 & 
--- & --- & --- & --- & 0.247 \\

$\gamma_{11}$ & 7.2 & -4.7 & 0.50 & 0.853 & 1.045 & 0.902 & 0.883 
& 0.906 & 0.915 & --- & 0.924 & 0.927 \\

$\gamma_{1}$ & 8.0 & -3.2 & 1.00 & 1.441 & 1.789 & 1.496 & 
1.432 & 1.501 & 1.512 & --- & 1.532 & 1.521 \\

$\delta_{1}$ & -4.3 & -1.7 & 5.00 & 6.764 & 7.725 & 6.762 & 5.861 & 
6.454 & 6.487 & 6.479 & 6.596 & 7.147 \\

$\delta_{11}$ & -7.8 & -1.7 & 3.00 & 4.411 & 5.664 & 4.231 & 3.699 & 
4.262 & 4.303 & 4.271 & 4.413 & 4.747 \\
    
\end{tabular}
\end{center}
\end{table}

\begin{table}[htb]
\caption{Surface critical exponents of the special transition for $d=3$ 
up to two-loop order at the cubic fixed point (of order $p=2$) 
$u^{*}=0.525, v^{*}=1.146$ at $n=8$.}
\label{tab4}
\begin{center}
\begin{tabular}{rrrrrrrrrrrrr}
\hline
$ exp $~&~$\frac{O_{1}}{O_{2}}$~&~$\frac{O_{1i}}{O_{2i}}$~&~$[0/0]$~&~
$[1/0]$~&~$[0/1]$~&~$[2/0]$~&~$[0/2]$~&~$[11/1]$~&~$[1/11]$~&~$ R $~&~
$ R^{-1} $~&~$ f(\eta_{\perp},\nu,\eta) $ 
\\ \hline
$\eta_{\parallel}$ & -5.2 & 6.2 & 0.00 & -0.355 & -0.262 & -0.286 & 
-0.292 & -0.304 & -0.292 & -0.308 & --- & -0.311 \\

$\Delta_{1}$ & 6.5 & -5.0 & 0.75 & 1.105 & 1.300 & 1.160 & 1.146 & 1.164
 & 1.178 & --- & 1.187 & 1.187 \\

$\eta_{\perp}$ & -3.8 & -11.6 & 0.00 & -0.177 & -0.151 & -0.131 & -0.140 & 
-0.145 & -0.142 & -0.148 & -0.143 & -0.143 \\

$\beta_{1}$ & 0.0 & 0.0 & 0.25 & 0.25 & 0.25 & 0.257 & 0.257 & 
--- & --- & --- & --- & 0.247 \\

$\gamma_{11}$ & 6.4 & -5.1 & 0.50 & 0.855 & 1.050 & 0.911 & 0.898 
& 0.913 & 0.931 & --- & 0.940 & 0.940 \\

$\gamma_{1}$ & 7.1 & -3.3 & 1.00 & 1.444 & 1.798 & 1.506 & 
1.447 & 1.509 & 1.532 & --- & 1.552 & 1.537 \\

$\delta_{1}$ & -4.8 & -1.8 & 5.00 & 6.775 & 7.751 & 7.125 & 5.920 & 
6.505 & 6.570 & 6.528 & 6.676 & 7.221 \\

$\delta_{11}$ & -9.6 & -1.7 & 3.00 & 4.420 & 5.695 & 4.272 & 3.750 & 
4.302 & 4.382 & 4.310 & 4.490 & 4.806 \\
    
\end{tabular}
\end{center}
\end{table}

\begin{table}[htb]
\caption{Surface critical exponents of the special transition for $d=3$ 
up to two-loop order at the cubic fixed point (of order $p=2$) 
$u^{*}=0.201, v^{*}=1.508$ at $n\to \infty$.}
\label{tab5}
\begin{center}
\begin{tabular}{rrrrrrrrrrrrr}
\hline
$ exp $~&~$\frac{O_{1}}{O_{2}}$~&~$\frac{O_{1i}}{O_{2i}}$~&~$[0/0]$~&~
$[1/0]$~&~$[0/1]$~&~$[2/0]$~&~$[0/2]$~&~$[11/1]$~&~$[1/11]$~&~$ R $~&~
$ R^{-1} $~&~$ f(\eta_{\perp},\nu,\eta) $ 
\\ \hline
$\eta_{\parallel}$ & -5.0 & 6.5 & 0.00 & -0.352 & -0.260 & -0.282 & 
-0.289 & -0.301 & -0.288 & -0.306 & --- & -0.307 \\

$\Delta_{1}$ & 6.7 & -4.9 & 0.75 & 1.102 & 1.293 & 1.154 & 1.140 & 1.157
 & 1.173 & --- & 1.182 & 1.182 \\

$\eta_{\perp}$ & -3.7 & -10.7 & 0.00 & -0.176 & -0.150 & -0.129 & -0.138 & 
-0.144 & -0.140 & -0.147 & -0.141 & -0.141 \\

$\beta_{1}$ & 0.0 & 0.0 & 0.25 & 0.25 & 0.25 & 0.258 & 0.258 & 
--- & --- & --- & --- & 0.248 \\

$\gamma_{11}$ & 6.6 & -5.0 & 0.50 & 0.852 & 1.043 & 0.905 & 0.891 
& 0.906 & 0.926 & --- & 0.935 & 0.935 \\

$\gamma_{1}$ & 7.4 & -3.3 & 1.00 & 1.440 & 1.785 & 1.499 & 
1.440 & 1.501 & 1.526 & --- & 1.546 & 1.531 \\

$\delta_{1}$ & -4.7 & -1.8 & 5.00 & 6.759 & 7.714 & 7.204 & 5.906 & 
6.492 & 6.564 & 6.515 & 6.669 & 7.179 \\

$\delta_{11}$ & -9.1 & -1.7 & 3.00 & 4.407 & 5.651 & 4.252 & 3.738 & 
4.287 & 4.376 & 4.296 & 4.483 & 4.772 \\
    
\end{tabular}
\end{center}
\end{table}

\begin{table}[htb]
\caption{Surface critical exponents involving the RG function 
$\eta_{\bar{c}}$ for the case $n=3,d=3$ at the cubic fixed point 
(of order $p=3$) $u^{*}=1.348, v^{*}=0.074$.}
\label{tab6} \begin{center}
\begin{tabular}{rrrrrrrrrrrrr}
\hline
$ exp $~&~
$\frac{O_{1}}{O_{2}}$~&~$\frac{O_{1i}}{O_{2i}}$~&~$[0/0]$~&~
$[1/0]$~&~$[0/1]$~&~$[2/0]$~&~$[0/2]$~&~$[11/1]$~&~$[1/11]$~&~$[R]$~&~
$R_{i}^{-1}$~&~$f(\alpha_{11},\nu,\eta)$
\\ 
\hline
$\eta_{\bar{c}}$ & -1.0 & -2.1 & 0.00 & -0.565 & -0.361 & 0.022 & -0.229 
 & -0.278 & -0.280 & -0.311 & -0.287 & -0.237 \\

$\alpha_{1}$ & -2.4 & 56.3 & 0.50 & 0.058 & 0.194 & 0.246 & 0.190 & 0.190 
& 0.188 & 0.176 & --- & 0.138 \\

$\alpha_{11}$ & -1.4 & -6.4 & 0.00 & -0.565 & -0.361 & -0.158 & -0.323 & 
-0.329 & -0.331 & -0.356 & -0.331 & -0.331\\

$\Phi$ & -0.6 & -0.6 & 0.50 & 0.377 & 0.390 & 0.597 & 0.589 & 0.455 & 0.455 
& 0.446 & 0.447 & 0.532\\

\end{tabular}
\end{center}
\end{table}

\begin{table}[htb]
\caption{Surface critical exponents involving the RG function 
$\eta_{\bar{c}}$ for the case $n=4,d=3$ at the cubic fixed point 
(of order $p=2$) $u^{*}=1.064, v^{*}=0.520$.}
\label{tab7} \begin{center}
\begin{tabular}{rrrrrrrrrrrrr}
\hline
$ exp $~&~
$\frac{O_{1}}{O_{2}}$~&~$\frac{O_{1i}}{O_{2i}}$~&~$[0/0]$~&~
$[1/0]$~&~$[0/1]$~&~$[2/0]$~&~$[0/2]$~&~$[11/1]$~&~$[1/11]$~&~$[R]$~&~
$R_{i}^{-1}$~&~$f(\alpha_{11},\nu,\eta)$
\\ 
\hline
$\eta_{\bar{c}}$ & -1.0 & -2.4 & 0.00 & -0.625 & -0.385 & 0.023 & -0.269 
 & -0.315 & -0.329 & -0.362 & -0.323 & -0.263 \\

$\alpha_{1}$ & -2.6 & 9.7 & 0.50 & 0.011 & 0.172 & 0.200 & 0.150 & 0.150 
& 0.137 & 0.125 & --- & 0.102 \\

$\alpha_{11}$ & -1.5 & -17.2 & 0.00 & -0.625 & -0.385 & -0.198 & -0.371 & 
-0.373 & -0.392 & -0.417 & -0.374 & -0.374\\

$\Phi$ & -0.6 & -0.6 & 0.50 & 0.364 & 0.380 & 0.602 & 0.591 & 0.443 & 0.442 
& 0.434 & 0.436 & 0.524\\

\end{tabular}
\end{center}
\end{table}

\begin{table}[htb]
\caption{Surface critical exponents involving the RG function 
$\eta_{\bar{c}}$ for the case $n=8,d=3$ at the cubic fixed point 
(of order $p=2$) $u^{*}=0.525, v^{*}=1.146$.}
\label{tab8} \begin{center}
\begin{tabular}{rrrrrrrrrrrrr}
\hline
$ exp $~&~
$\frac{O_{1}}{O_{2}}$~&~$\frac{O_{1i}}{O_{2i}}$~&~$[0/0]$~&~
$[1/0]$~&~$[0/1]$~&~$[2/0]$~&~$[0/2]$~&~$[11/1]$~&~$[1/11]$~&~$[R]$~&~
$R_{i}^{-1}$~&~$f(\alpha_{11},\nu,\eta)$
\\ 
\hline
$\eta_{\bar{c}}$ & -1.0 & -2.7 & 0.00 & -0.629 & -0.386 & -0.001 & -0.284 
 & -0.328 & -0.354 & -0.384 & -0.335 & -0.268 \\

$\alpha_{1}$ & -2.8 & 7.4 & 0.50 & 0.008 & 0.170 & 0.183 & 0.141 & 0.143 
& 0.120 & 0.109 & --- & 0.091 \\

$\alpha_{11}$ & -1.6 & -72.9 & 0.00 & -0.629 & -0.386 & -0.225 & -0.383 & 
-0.384 & -0.419 & -0.442 & -0.384 & -0.384 \\

$\Phi$ & -0.6 & -0.7 & 0.50 & 0.363 & 0.379 & 0.592 & 0.579 & 0.435 & 0.434 
& 0.427 & 0.429 & 0.525 \\

\end{tabular}
\end{center}
\end{table}

\begin{table}[htb]
\caption{Surface critical exponents involving the RG function 
$\eta_{\bar{c}}$ for the case $n\to\infty,d=3$ at the cubic fixed point 
(of order $p=2$) $u^{*}=0.201,  v^{*}=1.508 $.}
\label{tab9} \begin{center}
\begin{tabular}{rrrrrrrrrrrrr}
\hline
$ exp $~&~
$\frac{O_{1}}{O_{2}}$~&~$\frac{O_{1i}}{O_{2i}}$~&~$[0/0]$~&~
$[1/0]$~&~$[0/1]$~&~$[2/0]$~&~$[0/2]$~&~$[11/1]$~&~$[1/11]$~&~$[R]$~&~
$R_{i}^{-1}$~&~$f(\alpha_{11},\nu,\eta)$
\\ 
\hline
$\eta_{\bar{c}}$ & -1.0 & -2.7 & 0.00 & -0.624 & -0.384 & 0.001 & -0.279 
 & -0.327 & -0.356 & -0.385 & -0.333 & -0.266 \\

$\alpha_{1}$ & -2.8 & 8.0 & 0.50 & 0.012 & 0.172 & 0.189 & 0.146 & 0.148 
& 0.123 & 0.111 & --- & 0.095 \\

$\alpha_{11}$ & -1.5 & -38.2 & 0.00 & -0.624 & -0.384 & -0.218 & -0.378 & 
-0.380 & -0.419 & -0.441 & -0.380 & -0.380 \\

$\Phi$ & -0.6 & -0.7 & 0.50 & 0.364 & 0.380 & 0.593 & 0.580 & 0.434 & 0.433 
& 0.426 & 0.429 & 0.525 \\

\end{tabular}
\end{center}
\end{table}

\begin{table}[htb]
\caption{Surface critical exponents calculated on the basis of 
$\alpha_{11}$ for $n=3,4,8$ and $n\to \infty$ (see Tables 6-9) and 
correspondent six-loop perturbation theory results for bulk critical 
exponent of the correlation length $\nu$.} 
\label{tab11} 
\begin{center} 
\begin{tabular}{rrrrr} \hline
$ exp $~&~$ n=3 $~&~$ n=4 $~&~$ n=8 $~&~$n \to \infty $ \\ \hline

$\alpha_{11}$ & -0.331 & -0.374 & -0.384 & -0.380 \\

$\eta_{\bar{c}}$ & -0.234 & -0.262 & -0.270 & -0.268 \\

$\alpha_{1}$ & 0.129 & 0.099 & 0.096 & 0.102 \\

$\Phi$ & 0.541 & 0.527 & 0.520 & 0.518 \\

\end{tabular}
\end{center}
\end{table}


\begin{thebibliography}{10}
\bibitem{mit3}
This is an intuitive picture conveyed from the theory of bulk 
regular systems. Much less is known about the large-order behavior of 
perturbative expansions pertaining to infinite random systems (see Refs.
  \cite{B87,Mc94,AMR00}), especially at large space dimensionalities. 
At the present time there are no explicit results on large orders for surface
quantities, even in the absence of disorder.
\bibitem{mitSh88} We applied the formulas of $\beta$ functions, presented 
in \cite{Sh88} for the case of cubic anisotropic model with $m=1$ and 
$n=3$.
\bibitem{mit2} The values of $\nu$ and $\eta$ at $n=4,8$ and $n\to\infty$
are calculated from formulas, presented in \cite{Sh88} for the case of cubic 
anisotropic model.
\bibitem{mit5} In order to evaluate the difference between surface critical 
exponents of semi-infinite model with cubic anisotropy and surface critical 
exponents of standard semi-infinite $n$-component model 
\cite{DD81,DD83,DSh94,DSh98} we performed additional calculation for surface 
critical exponents on the basis of formulas from \cite{DSh94,DSh98} in the 
case of 3D semi-infinite model with $n=3$.
\bibitem{mit4} The investigation of the surface 
critical behavior of random semi-infinite systems at the special and 
ordinary surface transition was performed early in serial works 
\cite{UShH00,UH02}. The obtained results indicate that the surface critical 
exponents of random semi-infinite systems belong to the universality class 
of random fixed point.
\end{thebibliography}
\end{document}